\documentclass[article]{JHEP3}

\usepackage{amsmath,amssymb}
\usepackage[dvips]{graphics}
\usepackage{epsfig}

\newcommand{\be}{\begin{equation}}
\newcommand{\ee}{\end{equation}}
\newcommand{\bea}{\begin{eqnarray}}
\newcommand{\eea}{\end{eqnarray}}
\newcommand{\bean}{\begin{eqnarray*}}
\newcommand{\eean}{\end{eqnarray*}}
\newcommand{\beq}{\begin{eqnarray}}
\newcommand{\eeq}{\end{eqnarray}}
\newcommand{\nn}{\nonumber}

\relax

\def\d{\partial}

\def\a{\alpha'}
\def\R{{\mathcal R}}

\title{Absorption of scalars by extremal black holes in string theory}

\author{Filipe Moura
\\
LIP - Laborat\'orio de Instrumenta\c c\~ao e F\'\i sica Experimental de Part\'\i culas, \\Department of Physics, University of Coimbra\\ 3004-516 Coimbra, Portugal\\
\\
\email{fmoura@lip.pt}
}

\abstract{We show that the low frequency absorption cross section of minimally coupled test massless scalar fields by extremal spherically symmetric black holes in $d$ dimensions is equal to the horizon area, even in the presence of string--theoretical $\a$ corrections. Classically one has the relation $\sigma = 4 G S$ between that absorption cross section and the black hole entropy. By comparing in each case the values of the horizon area and Wald's entropy, we discuss the validity of such relation in the presence of higher derivative corrections for extremal black holes in many different contexts: in the presence of electric and magnetic charges; for nonsupersymmetric and supersymmetric black holes; in $d=4$ and $d=5$ dimensions. The examples we consider seem to indicate that this relation is not verified in the presence of $\a$ corrections in general, although being valid in some specific cases (electrically charged maximally supersymmetric black holes in $d=5$). We argue that the relation $\sigma = 4 G S$ should in general be valid for the absorption cross section of scalar fields which, although being independent from the black hole solution, have their origin from string theory, and therefore are not minimally coupled.
}

\begin{document}



\vfill

\eject

\section{Introduction and summary}
\label{int}
\indent

A classical result in black hole scattering is that in the low frequency limit, the absorption cross section of minimally coupled test \footnote{By \emph{test} fields with respect to a specific black hole solution, we mean fields which are not part of that solution. By \emph{minimally coupled} fields, we mean fields which obey the simplest possible second order free field equation on the background of such black hole solution. In the case of scalar fields, this means the Klein-Gordon equation in curved space.} scalar fields by arbitrary spherically symmetric black holes is equal to the horizon area. This has been shown first to some classes of five--dimensional extremal black holes \cite{Dhar:1996vu, Das:1996wn}, and soon after to general $d-$dimensional spherically symmetric black holes \cite{dgm96}. Also in \cite{dgm96} this result has been generalized to higher spin fields, but always in Einstein gravity.

It is interesting to study the effects of string $\a$ corrections in this quantity. This work has been started, concerning a particular black hole solution, in \cite{Moura:2006pz}. More recently, we have studied scattering of $\a$--corrected spherically symmetric nonextremal black holes in $d$ spacetime dimensions. We have shown \cite{Moura:2011rr, Moura:2014epa} that the $\a$--corrected absorption cross section $\sigma$ could be expressed solely in terms of the black hole horizon area $A_H$ and temperature $T$, without explicit $\a$ corrections: those leading corrections would be just implicit in $A_H$ and $T.$ Although this expression for the $\a$--corrected absorption cross section has a smooth behavior in the extremal limit, as we have shown in \cite{Moura:2014epa}, clearly it cannot be valid for extremal black holes, for which $T=0$.

In this article, we are therefore concerned with how are such $\a$ corrections to the low frequency absorption cross section expressed for extremal black holes.

The $\a$-corrected absorption cross section is known at least for a class of 1/4 BPS $\mathcal{N}=4$ black holes in string theory with leading $\a$ corrections both in $d=4$ and $d=5$ \cite{Kuperstein:2010ka}. In both cases the low frequency absorption cross section and the black hole entropy were computed, and the agreement $\sigma= 4 G S$ was obtained to first order in $\a,$ i.e. the leading $\a$ corrections for these two quantities were equal. This motivates us to investigate if this agreement is also verified for general extremal black holes, or at least to supersymmetric black holes, or if it is just a coincidence valid just for that particular class of 1/4 BPS $\mathcal{N}=4$ black holes.

The article is organized as follows. In section \ref{nhg}, by considering the near--horizon geometry of $\a$-corrected spherically symmetric extremal black holes, we obtain a formula for their low frequency absorption cross section. We keep the requirement of minimally coupling of the scalar field and of spherical symmetry, because only for these black holes general formulas for the absorption cross section are known. In section \ref{walds}, we briefly review Wald's formula and Sen's formalism to compute the entropy of $\a$-corrected extremal black holes. We then apply these results to different black hole solutions with higher derivative corrections. In section \ref{emdgb}, we consider black holes in Einstein-Maxwell-dilaton theory with Gauss-Bonnet corrections (these solutions are not stringy in general). In section \ref{itz}, we take the extremal Reissner--Nordstr\"om solution in $d=4$ with leading string $\a$ corrections. In section \ref{glc}, we take black holes in $d=4$ $\mathcal{N}=2$ supergravity, with full supersymmetric $\a$ corrections. In section \ref{senglc}, we consider dyonic black holes in heterotic string theory, also in $d=4$, just with Gauss-Bonnet (non-supersymmetric) corrections. In section \ref{d5}, we consider different kinds of charged black holes in $d=5$ supergravity with leading $\a$ corrections. Finally, in section \ref{murthy} we review the 1/4 BPS $\mathcal{N}=4$ black holes with $\a$ corrections for which $\sigma= 4 G S$ considered in \cite{Kuperstein:2010ka} to which we referred in the previous paragraph. In this case the absorption cross section was computed through a different method and corresponds to a different case: the corresponding scalar field is not external, but intrinsic to string theory. For each case we compute the $\a$-corrected absorption cross section and entropy, comparing the two quantities in order to verify if there exists any relation between them. We conclude by discussing and comparing our results considering different criteria in section \ref{discussion}.

\section{Near--horizon geometry of $\a$-corrected extremal black holes and the absorption cross section of minimally coupled scalar fields}
\label{nhg}
\indent

A generic spherically symmetric $d-$dimensional black hole metric can be written in the form
\be \label{schwarz}
ds^2=- f(r) dt^2+ \frac{d r^2}{g(r)}+ r^2 d \Omega^2_{d-2}.
\ee

As shown for instance in \cite{Sen:2005wa, Sen:2007qy}, generically a spherically symmetric $d-$dimensional extremal black hole, close to its horizon, has a $SO(2,1) \otimes SO(d-1)$ symmetry. This symmetry implies that near the horizon the metric is of the form $AdS_2 \otimes S^{d-2}$. The only $SO(2,1)\otimes SO(d-1)$ invariant forms on $AdS_2\otimes S^{d-2}$ are the 2-form corresponding to the volume form on $AdS_2$ and the $(d-2)$-form corresponding to the
volume form on $S^{d-2}$. This implies that any other $k$-form field for $k$ different than 1, 2, $d-3$ or $d-2$ will vanish in this background. Assuming that the lagrangian depends on the gauge fields through their field strengths, this means that, besides the metric, the relevant fields which can take non-trivial expectation values in this region are scalars $\{\phi_s\}$, purely electric 1-form gauge fields $A^i_\mu$ (with associated 2-form field strength $F^i_{\mu \nu}=\partial_{[\mu} A^i_{\nu]}$) and purely magnetic $(d-3)$-form gauge fields $B^I_{\mu_1\ldots \mu_{d-3}}$ (with associated $(d-2)$-form field strength $H^I_{\mu_1 \ldots \mu_{d-2}}= \partial_{[\mu_1} B^I_{\mu_2 \cdots \mu_{d-2}]}$). The most general background consistent with the $SO(2,1)\otimes SO(d-1)$ symmetry of the background geometry is therefore of the form
\bea \label{sen}
&& ds^2 \equiv g_{\mu\nu}dx^\mu dx^\nu = v_1 \left(-\rho^2 \ d\ \tau^2  + \frac{d\ \rho^2}{\rho^2} \right) + v_2 d\Omega^2_{d-2}, \\
&& \phi_s =u_s, \qquad F^{a}_{rt} = e^a, \qquad H^{a}_{l_1\cdots l_{d-2}} =
\frac{p^a}{\Omega_{d-2}} \, \epsilon_{l_1\cdots l_{d-2}}\, \sqrt{\det h}\, \label{senforms}
\eea
where $v_1$, $v_2$, $\{u_s\}$, the electric field strengths $\{e^a\}$ and the magnetic charges $\{p^a\}$ are constants parameterizing the background and $d\Omega_{d-2}^2\equiv h_{ij} d\theta^i d\theta^j$ denotes the line element on the unit $(d-2)$-sphere.

The simplest example is the one of a $d-$dimensional Reissner-Nordstr\"om-like extremal black hole, with a metric of the form (\ref{schwarz}) for which one has
\be
f(r)=g(r)=f_{ext}(r) \equiv \left(1-\left(\frac{R_H}{r}\right)^{d-3}\right)^2, \label{fext}
\ee
$R_H$ being the horizon radius. Introducing the change of coordinates
\be
\rho = r-R_H, \tau = \frac{(d-3)^2 t}{R_H^2}, \label{rho}
\ee
near $\rho=0$ we find
$f_{ext}(\rho) \simeq \left(\frac{(d-3) \rho}{R_H}\right)^2 + {\mathcal{O}} \left( \rho \right).$ This way, near the horizon this metric can be cast in the form (\ref{sen}) - the so-called Bertotti-Robinson metric, with $v_1=\left(\frac{R_H}{d-3}\right)^2$ and $v_2=R_H^2.$ From this change of coordinates we see clearly that the horizon of a metric of the form (\ref{sen}) is a $(d-2)-$sphere of radius $R_H = \sqrt{v_2}$ and area $A_H= v_2^{\frac{d-2}{2}} \Omega_{d-2},$ located at $\rho=0.$ Since this metric (\ref{sen}) is the near--horizon geometry of a spherically symmetric solution, from \cite{dgm96} this area is also equal to the low frequency absorption cross section.

As also argued in \cite{Sen:2005wa, Sen:2007qy, Kunduri:2007vf}, the $SO(2,1) \otimes SO(d-1)$ symmetry is preserved in the presence of $\a$ corrections: it is valid in any generally covariant theory of gravity coupled to matter fields. Therefore, spherically symmetric $\a$--corrected extremal black holes near the horizon also have the form (\ref{sen}), but with $\a$--dependent coefficients $v_1, v_2.$ This allows us to find the answer to the question we raised.

Scattering cross sections can be computed by separately solving the scalar field equation, which for minimally coupled fields is a second order differential equation, in different regions of the radial parameter: near the black hole horizon, at asymptotic infinity and in the intermediate region. By matching the corresponding solutions, one is able to determine the cross section. This technique was initiated in four dimensional Einstein gravity in \cite{u76}, and later generalized to arbitrary $d$ dimensions in \cite{Harmark:2007jy}. In \cite{Moura:2006pz, Moura:2011rr} this technique was extended to include string $\a$ corrections. 

As argued in \cite{Harmark:2007jy}, the absorption cross section can only be defined for asymptotically flat spacetimes, since only in this case one can have incoming and outgoing asymptotic particle states, and thus an $S$--matrix. For a spacetime which is asymptotically either de Sitter or anti–-de Sitter, there is no good notion of an $S$--matrix and thus an absorption cross section cannot be defined. Since for an asymptotically flat spacetime obviously the curvature corrections vanish asymptotically, they have no effect on the analysis of scattering: black hole solutions do not get $\a$ corrections in this region. This means that the effect of those curvature corrections, if it exists, comes only from the near--horizon region (the intermediate region is considered only for the purpose of matching). But the same is valid already at the classical level, as we mentioned \cite{u76,Harmark:2007jy}. This way we conclude that in general the low frequency absorption cross section depends only on the near--horizon geometry, even in the presence of $\a$ corrections.

But, as we have seen, for an extremal black hole (and differently than for the non--extremal case analyzed in \cite{Moura:2011rr}) the near horizon geometry is preserved by $\a$ corrections; for a spherically symmetric solution it is of the form (\ref{sen}). The calculations and the result for the low frequency absorption cross section for minimally coupled test scalar fields are therefore exactly the same as in the classical case: for an extremal black hole, it is simply given by the $\a$--corrected black hole horizon area,
\be
\sigma= A_H. \label{seccaoext}
\ee

As we mentioned, there is no explicit $\a$ dependence in the absorption cross section for non--extremal black holes. The same can be said for extremal black holes: $\a$ corrections to $\sigma$ in (\ref{seccaoext}) come only from the corresponding (implicit) $\a$ corrections to $A_H.$ Equation(\ref{seccaoext}) is valid classically (in the $\a \rightarrow 0$ limit), but also in the presence of $\a$ corrections, to all orders in $\a$ (in the same way as (\ref{sen})).

\section{Entropy of $\a$-corrected extremal spherically symmetric black holes}
\label{walds}
\indent

The entropy of a black hole in a $d$--dimensional theory with higher derivative terms, like $\a$ corrections, is given by Wald's formula
\be
S=-2 \pi \int_H \frac{\partial
\mathcal{L}}{\partial \R^{\mu \nu \rho \sigma}}
\varepsilon^{\mu\nu} \varepsilon^{\rho\sigma} \, \sqrt{h} \,
d\,\Omega_{d-2}, \label{wald0}
\ee
$\mathcal{L}$ being the full $d$--dimensional lagrangian, including $\a$ corrections, one is considering. $H$ is the black hole horizon, with area $A_H$ and metric $h_{ij}$ induced by the spacetime metric $g_{\mu\nu};$ $\varepsilon^{\mu\nu}$ is the binormal to $H.$ In order to consider the effect of the $\a$ corrections, we split the lagrangian into its purely gravitational classical and $\a$ corrected parts, and also the nongravitational part:
\be \label{eef}
\mathcal{L} = \frac{1}{16 \pi G} \sqrt{-g} \left( \R + \sum_i \a^i\ Y_i(\R) + \mathcal{L}_{\mathrm{matter}}\right).
\ee
\noindent
Here, $Y_i(\R)$ are higher order terms depending on the Riemann tensor, each of them multiplied by the suitable $i-$th power of the inverse string tension $\a$. These terms include (but are not restricted to) purely gravitational corrections, given by scalar polynomials in the Riemann tensor, eventually multiplied by dilaton terms. $\mathcal{L}_{\mathrm{matter}}$ contains classical and $\a$--correction terms depending on other matter fields and obviously also on the metric, but not on the Riemann tensor. When dealing with a black hole solution one considers a perturbative expansion in $\a$, and therefore one only takes in (\ref{eef}) the terms up to the order in $\a$ we are considering.

For the metric (\ref{schwarz}), the nonzero components of $\varepsilon^{\mu\nu}$ are $\varepsilon^{tr}=-\varepsilon^{rt}=-1.$ We have
$$8 \pi G\frac{\partial \mathcal{L}}{\partial \R^{\mu \nu \rho
\sigma}}=\frac{1}{4}\left(g_{\mu\rho}g_{\sigma\nu}-g_{\mu\sigma}g_{\rho\nu}\right)+
\sum_i \frac{\a^i}{2} \frac{\partial
Y_i(\R)}{\partial \R^{\mu \nu \rho \sigma}}.$$
This way, taking only nonzero components, one gets from (\ref{schwarz})
\be
8 \pi G \frac{\partial \mathcal{L}}{\partial
\R^{\mu \nu \rho \sigma}} \varepsilon^{\mu\nu}
\varepsilon^{\rho\sigma} = 4 \times 8 \pi G \frac{\partial
\mathcal{L}}{\partial \R^{trtr}} \varepsilon^{tr} \varepsilon^{tr}
=-1+ 2 \sum_i \a^i \frac{\partial
Y_i(\R)}{\partial \R^{trtr}}, \label{wald}
\ee
and therefore we obtain Wald's entropy formula for generic spherically symmetric black holes:
\be
S=\frac{1}{4 G} \int_H \left( 1 -2 \sum_i \a^i \frac{\partial
Y_i(\R)}{\partial \R^{trtr}} \right) \, \sqrt{h} \, d\,\Omega_{d-2} =\frac{A_H}{4 G}
- \sum_i \frac{\a^i}{2 G} \int_H \frac{\partial
Y_i(\R)}{\partial \R^{trtr}} \, \sqrt{h} \, d\,\Omega_{d-2}. \label{wald2}
\ee
Generically, (\ref{wald2}) can be written in the form
\be
S=\frac{A_H}{4 G} \left(1+ \sum_i \a^i \delta_i S \right), \label{entropy}
\ee
$\delta_i S$ corresponding to the specific correction entropy correction of order $\a^i.$ The $\a$ corrections to the entropy emerge as multiplicative factors to the classical formula.

In general, because of the $\a$ corrections, (\ref{wald2}) may be very difficult to compute directly. For extremal black holes having a near horizon $AdS_2$ factor (namely spherically symmetric black holes, as we have seen), Wald's entropy formula reduces to evaluating Sen's entropy function at its extremum \cite{Sen:2005wa}. This entropy function is given by the Legendre transformation of the action with respect to the electric charges (for a review of Sen's formalism see \cite{Sen:2007qy}).

We now consider different black hole solutions with higher derivative corrections. For each solution, we consider its near horizon geometry to obtain the low frequency absorption cross section of minimally coupled test scalar fields (which from (\ref{seccaoext}) is simply given by the horizon area), and we compare it to the entropy, in order to verify if the relation $\sigma= 4 G S$ holds.

\section{Einstein-Maxwell-dilaton-Gauss-Bonnet black holes}
\label{emdgb}
\indent

We start by considering Einstein-Maxwell-dilaton theory with a Gauss-Bonnet higher order correction in arbitrary $d$ dimensions:
\be \label{reducedS}
I = \frac1{16 \pi G} \int \sqrt{-g} \left[ \R - 2 \partial_\mu \phi \partial^\mu \phi
- {\rm e}^{2 a \phi} \left( F_{\mu\nu} F^{\mu\nu} - \alpha Y_{\rm GB} \right) \right]
\, d^dx,
\ee
$Y_{\rm GB}$ being the Gauss-Bonnet density
\be \label{gb}
Y_{\rm GB} = \R_{\mu\nu\rho\sigma} \R^{\mu\nu\rho\sigma}-4 \R_{\mu\nu} \R^{\mu\nu} +\R^2.
\ee
This action contains two parameters: the dilaton coupling constant $a$ and the Gauss-Bonnet coupling constant $\alpha$. The choices $a=\sqrt{\frac{2}{d-2}}, \alpha=\frac{\a}{8}$ corresponds to the values of these parameters as they would appear in an effective action from heterotic string theory (after rescaling the dilaton by a constant value), although such string action in this case would also have to contain other terms of order $\alpha$ including the dilaton and the Maxwell 2-form $F$. Here we leave these parameters as general and not necessarily perturbative. Different choices of the values of these parameters give rise to distinct physical properties; for such discussion we refer the reader to the original references. Therefore in general the solutions we take in this section do not have a stringy origin, but the same questions about entropy and cross section can be asked, and we think it is instructive to consider them.

\subsection{Four dimensional dyonic black holes}
\indent

We begin by taking $d=4$ in (\ref{reducedS}) and considering extremal dyonic black holes in this setting. The solutions for the metric, of the form (\ref{schwarz}), and for the Maxwell 1-form, were obtained in \cite{Chen:2006ge,Chen:2008hk}. The electric and magnetic charges $Q$ and $P$ respectively are independent quantities, in terms of which the horizon radius is fixed as
\begin{equation}
R_H^2 = \frac{2 Q (P^2 + 2\alpha)}{\sqrt{P^2 + 4 \alpha}}.
\end{equation}
At the horizon, the dilaton comes as $e^{2 a \phi(R_H)} = \frac{Q}{\sqrt{P^2 + 4 \alpha}}.$ Near this region the geometry is of the form (\ref{sen}), with $d=4$ and $v_1= v_2 = R_H^2.$ The low frequency cross section is $\sigma= 4 \pi  R_H^2$.

The black hole entropy was computed using Sen's entropy function method:
\be \label{entg}
S= \frac{\pi R_H^2}{G} \left(1 + \frac{2 \alpha}{2 \alpha + P^2} \right)= \frac{2\pi}{G} Q \sqrt{P^2 +4\alpha}.
\ee
From the first expression in (\ref{entg}), we see immediately that the corrected entropy is not equal to $\sigma/4G.$ The only possibility for that to happen is for $\alpha=0,$ i.e. setting the higher derivative corrections to 0. There is another limit in which something interesting happens -- setting the magnetic charge $P$ to 0, the corrected entropy becomes equal not to one quarter but to half of the horizon area: $S=\frac{\sigma}{2G}.$ In that limit, the black hole becomes small, as the horizon area goes to 0 in the limit $\alpha \rightarrow 0$.

\subsection{$d$-dimensional electrically charged black holes}
\indent

We now consider just purely electrical $d$-dimensional extremal black hole solutions of (\ref{reducedS}), with metric of the form (\ref{schwarz}) and a Maxwell 1-form with $P\equiv0$, which were obtained in \cite{Chen:2009rv}.

The near horizon geometry is of the form (\ref{sen}), with $v_1, v_2$ being functions of the electric charge $Q$ and $\alpha$:
\be
v_2^{d-2}=\left(\frac{16 \pi G}{\Omega_{d-2}}\right)^2 (2d-7)^2 \, \frac{2}{ (d-3)(d-2)\left(d^2-d-8 \right)} \, Q^2 \, \alpha, \, v_1 = \frac{2}{(d-3)(d-2)}v_2.
\ee
We see we are dealing with small black holes, since $v_1, v_2 \rightarrow 0$ when $\alpha \rightarrow 0$. The dilaton at the horizon is given by
\be
e^{2 a \phi(R_H)} = \frac{v_2}{4 (2d-7) \, \alpha}.
\ee
It is easy to see that the dilaton becomes singular at the horizon in the limit $\alpha \rightarrow 0$.

The black hole mass and the dilaton charge can also be expressed as functions of $Q, \alpha$: for details see \cite{Chen:2009rv}. The black hole entropy is again obtained using Sen's entropy function method, the result being
\be \label{entgd}
S= \left(1 + \frac{(d-2)(d-3)}{2(2d-7)} \right) \frac{A_H}{4G}.
\ee
For $d=4$ we recover $S=\frac{\sigma}{2G}$ as in (\ref{entg}). Also like in (\ref{entg}), in the absence of magnetic charges (i.e. in the small black hole limit) the quotient $G \, S/\sigma$ is a number, which in this case only depends on the spacetime dimension $d$.

The proportionality constant between $\sigma$ and $S$ being different than $4 G$ is related to the fact that we are dealing with small black holes. The relation $S= \frac{A_H}{4G}$ is valid for black holes in classical Einstein-Hilbert gravity. In this case, small black holes correspond to naked spacetime singularities: it only makes sense to talk about small black holes in the presence of higher order terms. We will discuss this topic later on section \ref{dischar}.

\section{Extremal Reissner-Nordstr\"om metric with leading $\a$ corrections in $d=4$}
\label{itz}
\indent

From now on the solutions we consider come in a string theory context. We start by the classical Reissner-Nordstr\"om solutions with $\a$ corrections.

We take the string effective action with purely gravitational $\a$ corrections, in $d$ dimensions and in the Einstein frame, given by
\be \label{itef} I= \frac{1}{16 \pi G} \int \sqrt{-g} \left( \R -
\frac{4}{d-2} \left( \d^\mu \phi \right) \d_\mu \phi +
\mbox{e}^{\frac{4}{d-2} \phi} \frac{\lambda}{2}\
\R^{\mu\nu\rho\sigma} \R_{\mu\nu\rho\sigma} \right) \mbox{d}^dx.
\ee
Here $\lambda = \frac{\a}{2}, \frac{\a}{4}$ and $0$, for bosonic, heterotic and type II strings, respectively. This effective action would also include other terms depending on gauge fields and fermionic fields, but these can be consistently set to 0, as they do not matter to the solution we are interested in. Besides the metric, only the dilaton field cannot be set to 0: such choice would be inconsistent with the Einstein and dilaton field equations, because of the dilaton dependence of the $\lambda$ term in (\ref{itef}).

Taking $d=5$ and expanding this action around the vacuum $M^4 \otimes S^1,$ $S^1$ being a circle of radius $R$, one obtains a five dimensional line element given by
\be \label{4d5} ds^2_5=e^{4\sqrt{3}\kappa\sigma}(dx^5+2\kappa A_{\mu}dx^{\mu})^2+
e^{-2\sqrt{3}\kappa\sigma}g_{\mu\nu}dx^{\mu}dx^{\nu}, \ee
where $G=G_5/2\pi R$, $\kappa^2=4\pi G$, $A_{\mu}$ is an abelian gauge field, $\sigma$ is a scalar field and $g_{\mu\nu}$ is the four dimensional metric: a four dimensional black hole which solves the five dimensional graviton field equation from (\ref{itef}).

The solution we consider \cite{Itzhaki:1997fm} is a spherically symmetric metric of the form (\ref{schwarz}), with $d=4$ and $f(r), g(r)$ given to order $\lambda$ by
\be
f(r)=g(r)=1-\frac{2R_H}{r}+\frac{q(r)^2 + p(r)^2}{r^2} \label{solitz}
\ee
with
\bea
q(r)&=&\frac{R_H}{\sqrt{2}}- \frac{\lambda}{R_H^2} \frac{\sqrt{2}}{140}
\frac{r-R_H}{r^4} \left[364 \log \left(\frac{r}{r-R_H}\right)
r^2 (r-R_H)^2 \right. \nn \\
&-& \left. 122 R_H^4 - 142 R_H^3 r+545 R_H^2 r^2-421 R_H r^3 \right], \label{solitzq} \\
p(r)&=&\frac{R_H}{\sqrt{2}}. \label{solitzp}
\eea
The four dimensional metric (\ref{solitz}) is the extremal dyonic Reissner-Nordstr\"om metric with leading $\a$ corrections, and with horizon radius $R_H$. Together with this solution, we have \cite{Itzhaki:1997fm}
\bea
&&A_{\mu}dx^{\mu}=\frac{q(r)}{r}dt+p(r) \cos \theta d\phi, \label{solitza} \\
&& \sigma (r)=\frac{\lambda}{R_H^2} \frac{1}{105 \sqrt{3} r^4}
\left[ 1092 r^2 (r-R_H)^2 \log \left(\frac{r}{r-R_H} \right)+1204 r^3 R_H-1697 r^2 R_H^2 \right. \nn \\ &&\left.
+360 r R_H^3 +225 R_H^4 \right]. \label{solitzx}
\eea

At the horizon, we have $q(R_H)=p(R_H)=\frac{R_H}{\sqrt{2}}.$ Close to the horizon, from (\ref{solitz}), (\ref{solitzq}) and (\ref{solitzp}) we have $f(r) \simeq \left(1-\frac{R_H}{r}\right)^2 + {\mathcal{O}} \left( r-R_H \right).$ Introducing the change of coordinates (\ref{rho}) with $d=4,$ near $\rho=0$ we find $f(\rho) \simeq \left(\frac{\rho}{R_H}\right)^2 + {\mathcal{O}} \left( \rho \right).$ This way in this region the metric (\ref{solitz}) can be cast in the form (\ref{sen}), with $d=4$ and $v_1= v_2 = R_H^2.$

At infinity one finds the $\lambda$ corrected ADM mass $M$, electric charge $Q$ and magnetic charge $P$ \cite{Itzhaki:1997fm}:
\beq
&& M=R_H \left(1+ \frac{16}{35} \frac{\lambda}{R_H^2} \right),\\
&& Q=\lim_{r \rightarrow \infty} q(r) =\frac{R_H}{\sqrt{2}} \left(1 + \frac{57}{70} \frac{\lambda}{R_H^2} \right), \\
&& P=\lim_{r \rightarrow \infty} p(r) =\frac{R_H}{\sqrt{2}}.
\eeq
One can invert any of these relations and, solving for $R_H$ in terms of either $M, Q$ or $P,$ express the solutions (\ref{solitz}), (\ref{solitza}) and (\ref{solitzx}) as functions of any of these quantities. (Since the solution only has one free parameter $R_H$, only one of these quantities is independent.) One therefore cannot consider the small black hole limit for this solution. Similarly we can solve for the horizon area $A_H = 4 \pi R_H^2$ and, from (\ref{seccaoext}), obtain the absorption cross section in terms of either the black hole mass, electric or magnetic charge:
\be
\sigma = A_H = 4 \pi M^2 \left(1- \frac{32}{35} \frac{\lambda}{M^2} \right) = 8 \pi Q^2 \left(1- \frac{57}{35} \frac{\lambda}{Q^2} \right) = 8 \pi P^2. \label{seccaoitz}
\ee
We see that, if expressed in terms of the magnetic charge, the cross section has no explicit $\lambda$ corrections. This is because the magnetic charge (actually even the function $p(r)$) itself has no explicit $\lambda$ corrections.

From (\ref{itef}) the relevant higher order correction to order $\lambda$ is given by $\mbox{e}^{\frac{4}{d-2} \phi} \frac{\lambda}{2}\ \R^{\mu\nu\rho\sigma} \R_{\mu\nu\rho\sigma}$. One has $\frac{\partial}{\partial \R^{trtr}} \left(\R^{\mu\nu\rho\sigma} \R_{\mu\nu\rho\sigma}\right) =2 \R_{trtr},$ and for a metric of the form (\ref{schwarz}), $\R_{trtr} = \frac{1}{2} f''.$ At order $\lambda=0,$ one has $\phi=0$ and, from (\ref{solitz}), $f''(R_H)=\frac{2}{R_H^2}.$ Therefore from (\ref{wald2}) we obtain for the black hole entropy of this solution
\be
S=\frac{1}{4 G} \int_H \left( 1 - \lambda \ f''(R_H) \right) \, \sqrt{h} \, d\,\Omega_{d-2} = \frac{A_H}{4 G} \left(1- \frac{2 \lambda}{R_H^2} \right).
\label{entitz}
\ee
One can also easily express the correction term $- \frac{2 \lambda}{R_H^2}$ in terms of the black hole charges or mass. In any case this term is different from zero: the entropy of this solution is not equal to $\frac{A_H}{4 G},$ to first order in $\lambda.$

\section {Four dimensional black holes in $\mathcal{N}=2$ supergravity}
\label{glc}
\indent

We now consider black holes which, classically, are solutions of $d=4$ $\mathcal{N}=2$ supergravity coupled to $n_V$ vector multiplets, with leading higher derivative corrections.

The basic object of ${\cal N}=2$, $d=4$ supergravity is the chiral superfield ${\bf W}^{ab}_{mn}$, whose $\theta$ expansions contain the components of the Weyl multiplet. Its lowest $\theta$ component is a complex antiselfdual auxiliary tensor field $T^{ab}_{mn}$; its highest component contains the antiselfdual part of the Weyl tensor. \footnote{In $d=4$, the Weyl tensor
can still be decomposed in its selfdual and antiselfdual parts: ${\cal W}_{\mu \nu \rho \sigma}= {\cal W}^+_{\mu \nu \rho \sigma} +
{\cal W}^-_{\mu \nu \rho \sigma}, {\cal W}^{\mp}_{\mu \nu \rho
\sigma} :=\frac{1}{2} \left({\cal W}_{\mu \nu \rho \sigma} \pm
\frac{i}{2} \varepsilon_{\mu \nu}^{\ \ \ \lambda \tau} {\cal
W}_{\lambda \tau \rho \sigma} \right).$} From this superfield we construct another chiral superfield as ${\bf W}^2 = \varepsilon_{ac} \varepsilon_{bd} {\bf W}^{ab}_{mn} {\bf W}^{mn cd}$, whose lowest $\theta$ component is given by $\Upsilon := T^{mn ab} T^{cd}_{mn} \varepsilon_{ac} \varepsilon_{bd},$ and whose highest component is given at the linearized level by $C := 64 {\cal W}_-^{\mu \nu \rho \sigma} {\cal W}^-_{\mu \nu \rho \sigma}$.

Each abelian vector multiplet contains a scalar $X^I, \, I=0, \ldots, n_V$, a doublet of chiral fermions $\Omega^I_a,$ a vector gauge field $W_m^I$ with field strength $F^{I}_{mn}$ and a real SU(2) triplet of scalars $Y^I_{ab}$. Each field strength $F^{I}_{mn}$ can also be decomposed into its selfdual and antiselfdual parts, like the Weyl tensor.

The lagrangian of the theory is written in terms of a chiral background multiplet, whose bosonic components are a scalar $\hat A$, a symmetric SU(2) tensor $\hat
B_{ab}$, an antiselfdual Lorentz tensor $\hat F^-_{mn}$ and another scalar $\hat C$. It is given in terms of a function $\mathcal{F}(X^I, \hat A)$ of the scalars $X^I$ and the auxiliary variable $\hat A$ which is holomorphic and homogeneous of degree two. We define
$\mathcal{F}_I := \frac{\partial}{\partial X^I} \mathcal{F}(X^I,\hat A), \,
\mathcal{F}_{\hat A} := \frac{\partial}{\partial \hat A} \mathcal{F}(X^I,\hat A), \,
\mathcal{F}_{I_1 \cdots I_k \hat A \cdots \hat A} =
\frac{\partial}{\partial X^{I_1}} \cdots
\frac{\partial}{\partial X^{I_k}}
\frac{\partial}{\partial \hat A} \cdots
\frac{\partial}{\partial \hat A} \mathcal{F}(X^I,\hat A) \,.$

Explicitly, the Lagrangian is given by
\bea
\label{efflag}
\frac{8 \pi G}{\sqrt{-g}} {\cal L}_0 &=&  - \frac{1}{2} \, \R +
\Big[ i {\cal D}^{\mu} \mathcal{F}_I \, {\cal D}_{\mu} \bar X^I
-\frac{1}{8}i  \mathcal{F}_{IJ}\, Y^I_{ab} Y^{Jab} - \frac{1}{4} i \hat
B_{ab}\,\mathcal{F}_{{\hat A}I}  Y^{Iab} \nonumber\\
&&+\frac{1}{4} i \mathcal{F}_{IJ} (F^{-I}_{mn} -\frac{1}{4} \bar X^I
T_{mn}^{ab}\varepsilon_{ab})(F^{-J}_{mn} -\frac{1}{4} \bar X^J
T_{mn}^{ab}\varepsilon_{ab}) \nonumber\\
&&-\frac{1}{8} i \mathcal{F}_I(F^{+I}_{mn} -\frac{1}{4}  X^I
T_{mnab}\varepsilon^{ab}) T_{mnab}\varepsilon^{ab}
+\frac{1}{2} i \, \mathcal{F}_{{\hat A}I} (F^{-I}_{mn} - \frac{1}{4}  \bar X^I
T_{mn}^{ab}\varepsilon_{ab}) \nonumber \\
&&+\frac{1}{2} i \mathcal{F}_{\hat A}
\hat C -\frac{1}{8} i \mathcal{F}_{{\hat A}{\hat A}}(\varepsilon^{ac}
\varepsilon^{bd}  \hat B_{ab} \hat B_{cd} -2 \hat F^-_{mn}\hat F^-_{mn})
-\frac{1}{32} i\mathcal{F} (T_{mnab}\varepsilon^{ab})^2 + {\rm h.c.} \Big]
\;\;.
\eea

In order for it to correspond to a supersymmetric effective action with quadratic corrections in the Weyl tensor, we identify the chiral background multiplet with ${\bf W}^2.$ This means in particular we identify $\hat
A$ with $\Upsilon$, $\mathcal{F}_{\hat A}$ with $\mathcal{F}_{\Upsilon}$, $\hat C$ with $C$.

We assume the function $\mathcal{F}(X^I,\Upsilon)$ has a power expansion around $\Upsilon=0,$ therefore defining a family of functions $\mathcal{F}^{(g)}(X^I)$ by $\mathcal{F}\left(X^I,\Upsilon\right) = \sum_{g=0}^{\infty} \Upsilon^{g} \mathcal{F}^{(g)}\left(X^I\right) \;.$ The first function $\mathcal{F}^{(0)}\left(X^I\right)$ in the expansion is the prepotential, and gives the minimal terms in the action.

The black holes we now consider are static, spherically symmetric solitonic interpolations between two $\mathcal{N}=2$ supersymmetric groundstates: flat Minkowski spacetime at infinity and the Bertotti-Robinson spacetime (\ref{sen}) at the horizon, preserving $\mathcal{N}=1$ supersymmetry. They were studied in \cite{Lopes Cardoso:1998wt, Lopes Cardoso:1999ur}. The entropy of these black holes is given by Wald's formula as ($Z$ being the value of the central charge)

\be
S = \pi \left(G^{-1} |Z|^2 - 256 \;\mbox{Im}\; \mathcal{F}_{\Upsilon}(X,\Upsilon)\right) = \frac{A_H}{4G}- 256 \pi\;\mbox{Im}\; \mathcal{F}_{\Upsilon}(X,\Upsilon)
\;,\;\;\;
\mbox{where}\;\;\; \Upsilon = - 64 \;.
\label{ModIndEntropyFormula}
\ee
The $X^I$ and $Z$ are specified in terms of the electric and magnetic charges $(q_I, p^I)$ carried by the extremal black hole as
\bea
|Z|^2 &=& p^I \mathcal{F}_I (X,\Upsilon) - q_I X^I, \\
X^I - \bar X^I &=& i p^I, \\
\mathcal{F}_I (X,\Upsilon) - \overline{\mathcal{F}}_I \left(\bar X, \bar \Upsilon \right) &=& i q_I.
\eea

Near the horizon, the metric of these black holes is of the form (\ref{sen}), with $v_2= |Z|^2.$ We see that the first term of (\ref{ModIndEntropyFormula}) corresponds to 1/4 of the horizon area, and therefore 1/4 of the absorption cross section. Supersymmetric higher order terms in the action in general introduce corrections to the entropy that turn it different than 1/4 of the absorption cross section.

\section {Four dimensional dyonic black holes in heterotic string theory}
\label{senglc}
\indent

We now consider four dimensional black hole solutions of heterotic string theory compactified on $\mathcal{M}_4 \otimes S^1 \otimes \tilde{S}^1,$ $\mathcal{M}_4$ being some suitable compact four manifold (e.g. $T^4$ or $K3$) and $S^1, \tilde{S}^1$ being circles of radii $R, \tilde{R}$ associated to coordinates $x^9, x^8,$ respectively. These black holes are in the presence of antisymmetric tensor fields.

The relevant four dimensional fields for the solution we consider come from the ten dimensional string metric $G_{MN}^{(10)}$, anti-symmetric tensor field $B_{MN}^{(10)}$ and dilaton $\Phi^{(10)}$ through
\bea
g_{\mu\nu}&=& G^{(10)}_{\mu\nu} - (G^{(10)}_{99})^{-1} \,
G^{(10)}_{9\mu} \, G^{(10)}_{9\nu} - (G^{(10)}_{88})^{-1} \,
G^{(10)}_{8\mu} \, G^{(10)}_{8\nu}, \nonumber \\
\phi &=& \Phi^{(10)} - \frac{1}{4} \ln (G^{(10)}_{99})
 - \frac{1}{4} \ln (G^{(10)}_{88}) - \frac{1}{2} \ln V_{\cal M}\, ,
S = e^{-2\phi}\, , R  = \sqrt{G^{(10)}_{99}}\, , \tilde{R} = \sqrt{G^{(10)}_{88}}, \nonumber \\
A^{(1)}_\mu &=& \frac{1}{2} (G^{(10)}_{99})^{-1} \, G^{(10)}_{9\mu}\, , A^{(2)}_\mu = \frac{1}{2}
(G^{(10)}_{88})^{-1} \, G^{(10)}_{8\mu}\, , A^{(3)}_\mu = \frac{1}{2} B^{(10)}_{9\mu}\, , A^{(4)}_\mu = \frac{1}{2} B^{(10)}_{8\mu},
\eea
$V_{\cal M}$ being the volume of $\mathcal{M}_4$ measured in the string metric. To these fields corresponds the four dimensional classical action (in the string frame)
\begin{eqnarray}
I_{cl} &=& \frac{1}{ 16\pi G} \int \, \sqrt{-g} \, e^{-2\phi} \, \bigg[
\mathcal{R} + 4 g^{\mu\nu} \, \partial_\mu \phi \partial_\nu \phi - R^{-2} \, g^{\mu\nu} \, \partial_\mu R \partial_\nu R - \widetilde{R}^{-2}
\, g^{\mu\nu} \, \partial_\mu \widetilde{R} \partial_\nu \widetilde{R} \label{sencl} \\
&-& R^2 \, g^{\mu\nu} \, g^{\mu'\nu'} \, F^{(1)}_{\mu\mu'}
F^{(1)}_{\nu\nu'} - \widetilde{R}^2 \, g^{\mu\nu} \, g^{\mu'\nu'} \,
F^{(2)}_{\mu\mu'} F^{(2)}_{\nu\nu'} - R^{-2} \, g^{\mu\nu} \, g^{\mu'\nu'} \, F^{(3)}_{\mu\mu'}
F^{(3)}_{\nu\nu'} - \widetilde{R}^{-2} \, g^{\mu\nu} \, g^{\mu'\nu'} \, F^{(4)}_{\mu\mu'}
F^{(4)}_{\nu\nu'}\bigg] \, d^4 x. \nonumber
\end{eqnarray}

We denote by $n$ and $w$ the number of units of momentum and winding along $S^1$ and by $N$ and $W$ the number of units of Kaluza-Klein and $H$-monopole charges associated with $\tilde{S}^1$. (A Kaluza-Klein monopole associated with $\tilde{S}^1$ represents a background where this circle is non-trivially fibered over a two sphere labelled by
the angular coordinates $\theta,\phi$. An $H$-monopole associated with $\tilde{S}^1$ represents a five-brane wrapped on $\mathcal{M}_4 \otimes S^1$.) We could have also considered, besides these charges, also momentum and winding along $\tilde{S}^1$ and monopole charge associated with $S^1$, but we set these charges to 0.

From (\ref{sencl}), the fields $A^{(1)}_\mu$ and $A^{(3)}_\mu$ couple to the momentum and winding numbers along the $x^9$ direction, whereas the fields $A^{(2)}_\mu$ and $A^{(4)}_\mu$ couple to the momentum and winding numbers along the $x^8$ direction. Thus the electric charges associated with the fields $A^{(1)}_\mu$ and $A^{(3)}_\mu$ are respectively proportional to $n$ and $w$, while the magnetic charges associated with the fields $A^{(2)}_\mu$ and
$A^{(4)}_\mu$ are respectively proportional to $N$ and $W$.

The compactification we consider allows for $\mathcal{N}=4$ supersymmetry in $d=4$. We are interested in a black hole solution with leading $\a$ corrections. In order to obtain the correct black hole, in principle we should consider the full set of terms of order $\a$ in the effective action, but for $\mathcal{N}=4$ supergravity in $d=4$ that full set is not known. But, as it was shown in \cite{Sen:2005iz}, it turns out that the entropy of such black hole can be determined through Sen's formalism by considering only the gravitational corrections, in this case given by the Gauss--Bonnet combination. This way, to the lagrangian in $I_{cl}$ we add a correction term of the form (again in the string frame)
\be
\frac{\lambda}{32\pi\,G}\, e^{-2\phi} Y_{\rm GB}, \label{sengb}
\ee
$Y_{\rm GB}$ being given by (\ref{gb}) and $\lambda$ having the same meaning as in (\ref{itef}). Of course the resulting black hole configuration is not supersymmetric, since it results from a non-supersymmetric action, but we consider it nonetheless.

The near horizon geometry of this solution, associated to the classical action (\ref{sencl}) and the correction (\ref{sengb}), is of the form (\ref{sen}) in $d=4$, with \cite{Sen:2007qy, Sen:2005iz}
\be
v_1=v_2=4\,N\,W+\frac{4\,\lambda}{G}. \label{senv1v2}
\ee
We see that, in the absence of magnetic charges ($N, W=0$) the black holes become small.

The near-horizon configuration of the remaining fields is given by
\bea
e^{-2\,\phi(R_H)} &=& \sqrt{\frac{nw}{NW+2\frac{\lambda}{G}}},\, R=\sqrt{\frac{n}{w}}, \, \tilde{R} = \sqrt{\frac{W}{N}}, \label{sendil} \\
F^{(1)}_{rt} &=& e^{2\,\phi(R_H)} w, \, F^{(3)}_{rt}=e^{2\,\phi(R_H)} n, \, F^{(2)}_{\theta\phi}=N, \, F^{(4)}_{\theta\phi} = W.  \label{senf}
\eea
This configuration results, via Sen's formalism, in a black hole entropy (computed in the Einstein frame) given, to first order in $\lambda$, by
\be
S=\frac{4\,\pi}{G}\sqrt{n\,w\,\left(NW+2\,\frac{\lambda}{G}\right)} \approx \frac{4\,\pi}{G}\sqrt{n\,w} \sqrt{N\,W} \left(1+ \frac{\lambda}{GNW}\right).
\ee
Here we notice that the value of the entropy would be the same if it had been computed in the string frame, as it can be checked for instance in \cite{Sen:2007qy} where both calculations were presented and compared. This reflects the scale invariance of the entropy function, which is remnant of the invariance of the lagrangian under local scale transformations.

For the absorption cross section, given by the horizon area, we get (also in the Einstein frame), from (\ref{senv1v2}) and (\ref{sendil})
\be
\sigma=A_H=e^{-2\,\phi(R_H)} 4\,\pi\,v_2 \approx 16\,\pi\, \sqrt{n\,w} \sqrt{N\,W}.
\ee
Comparing the two expressions we see that, like in the previous section, we do not find agreement between the entropy of this solution and $\frac{A_H}{4 G},$ to first order in $\lambda.$ Like in section \ref{emdgb}, the proportionality between $S$ and $\sigma$ only exists in the small black hole limit, but with a proportionality factor different than $\frac{1}{4G}.$

\section{Charged black holes in $d=5$ supergravity}
\label{d5}
\indent

We now consider black holes which, classically, are solutions of $d=5$ $\mathcal{N}=2$ supergravity coupled to $n_V$ vector multiplets, with leading higher derivative corrections.

\subsection{$\mathcal{N}=2$, $d=5$ supergravity and higher order corrections}
\indent

The field content of the Weyl multiplet is given by the vielbein $e_\mu^{~m},$ the gravitini $\psi_\mu,$ the vector boson $V_\mu$ associated with the gauging of $SU(2)$ $R$-symmetry, the gauge field of dilatational symmetry $b_\mu$ (when reducing to Poincar\'e supergravity, these two gauge fields are gauged way), and three auxiliary fields: the antisymmetric tensor $v_{mn}$, the fermion $\chi$, and the scalar $D$. Each vector multiplet consists of the gauge field $A_\mu^I$, the scalar $M^I$, the gaugino $\Omega^I$, and a scalar $Y^{I}$, which will also be gauged away. The functions defining the scalar manifold are the prepotential $\cal M$, characterized by the numbers $c_{IJK}$, and its derivatives:
\be
{\cal M}=\frac{1}{6} c_{IJK} M^I M^J M^K~,\quad {\cal M}_I=\partial_I {\cal M}=\frac{1}{2} c_{IJK} M^J M^K~,\quad {\cal M}_{IJ}=c_{IJK} M^K~, \label{prep}
\ee
where $I,J,K= 1, \ldots, n_V$.

The bosonic terms of the corresponding classical lagrangian are given by
\bea
\frac{16 \pi G}{\sqrt{-g}} {\cal L}_0&=& -\frac{1}{2} D+ \frac{3}{4} \R+v^2+{\cal M} \left( \frac{1}{2}D +\frac{1}{4} \R +3v^2\right)+2{\cal M}_I v^{mn} F^I_{mn} \nonumber \\
&+& {\cal M}_{IJ}\left(\frac{1}{4} F^I_{mn} F^{Jmn}+\frac{1}{2} \partial_m M^I \partial^m M^J \right)-\frac{1}{24}
c_{IJK} A^I_{m} F^{J}_{np} F^{K}_{qr} \epsilon^{mnpqr}. \label{l0}
\eea
Once one integrates out the auxiliary fields $D$ and $v_{mn}$, ${\cal L}_0$ yields the familiar $\mathcal{N}=2$ lagrangian arising from the compactification of eleven-dimensional supergravity on a Calabi-Yau 3-fold with intersection numbers $c_{IJK}$ and whose cohomology is completely described by the two independent Hodge numbers $h^{1,1}$ and $h^{2,1}$. In this case, the number of vector multiplets is given by $n_V=h^{1,1}$; from this compactification, $h^{2,1}+1$ hypermultiplets also arise, but we are not considering them. The scalars $M^I$ arise as moduli (volumes of the (1,1) cycles). Every term in the lagrangian becomes proportional to $\mathcal{M},$ which is seen as the volume of the Calabi-Yau manifold. At the two-derivative level, the field equation for $D$ requires the constraint $\mathcal{M}=1,$ defining the scalar real special geometry.

The four-derivative lagrangian ${\cal L}_1$ representing the bosonic part of the supersymmetrization of the gauge-gravitational Chern-Simons term is given by
\bea
\frac{16 \pi G}{\sqrt{-g}} {\cal L}_1&=& \frac{c_{I}}{24} \left( -\frac{1}{16} \epsilon_{m n p q r } A^{Im} C^{n p s u}C^{q r }_{~~s u} + \frac{1}{8} M^I C^{m n p q } C_{m n p q } +\frac{1}{12} M^I D^2 +\frac{1}{6} F^{Im n } v_{m n } D \right. \nonumber \\
&+& \frac{1}{3} M^I C_{m n p q } v^{m n } v^{p q } + \frac{1}{2} F^{Im n } C_{m n p q } v^{p q } - \frac{8}{3} M^I v_{m n } \hat{\cal D}^n  \hat{\cal D}_p  v^{m p } \nonumber \\
&-& \frac{4}{3} M^I \left({\hat{\cal D}}^m  v^{n p }\right) {\hat{\cal D}}_m  v_{n p } - \frac{4}{3} M^I \left({\hat{\cal D}}^m  v^{n p }\right) {\hat{\cal D}}_n  v_{p m } + \frac{2}{3} M^I \epsilon_{m n p q r } v^{m n } v^{p q } {\hat{\cal D}}_s  v^{r s } \nonumber \\
&-& \frac{2}{3} F^{Im n } \epsilon_{m n p q r } v^{p s } {\hat{\cal D}}_s  v^{q r } - F^{Im n } \epsilon_{m n p q r } v^p _{~s } {\hat{\cal D}}^q  v^{r s } \nonumber \\
&-& \left. \frac{4}{3} F^{Im n } v_{m p } v^{p q } v_{q n } - \frac{1}{3} F^{Im n } v_{m n } v^2 +4 M^I v_{m n } v^{n p } v_{p q } v^{q m } - M^I (v^2)^2 \right)~. \label{l1cs}
\eea
$C_{\mu \nu \rho \sigma }$ is the Weyl tensor; one also defines ${\hat{\cal D}}_\mu = {\cal D}_\mu  - b_\mu$. All the terms in this lagrangian are summed over the vector multiplet indices $I$; each of them is multiplied by the respective coupling constant $c_{I}$ corresponding to the higher derivative corrections, like $\a$ in string theory. If one considers M-theory compactified on a Calabi-Yau manifold, ${\cal L}_1$ arises from a term predicted by M5-brane anomaly cancelation \cite{Ferrara:1996hh}. The set of topological coefficients $c_{I}$ depends on the manifold one takes, but each of the $c_{I}$ is proportional to its second Chern class. The field equation for $D$ is now
\be
{\cal M} = 1 - \frac{c_{I}}{72} \left( M^I D + F^{Im n } v_{m n }\right) \label{fed}
\ee
implying that the scalar real special geometry constraint $\mathcal{M}=1$ is no further valid, except for special cases. Notice that the functional form $\mathcal{M}=\frac{1}{6} c_{IJK} M^I M^J M^K$ from (\ref{prep}) is still valid in the presence of string $\a$ corrections, with the same intersection numbers $c_{IJK}$; only string loop (in $g_S$) corrections would bring also corrections to the coefficients $c_{IJK}$.

Like we did with the four dimensional solutions of the previous section, we are now going to compare the near--horizon geometry and absorption cross section for black holes which in this case correspond to solutions of the lagrangian given by ${\cal L} = {\cal L}_0 + {\cal L}_1$, ${\cal L}_0$ being the classical part (\ref{l0}) and ${\cal L}_1$ the higher order corrections (\ref{l1cs}).

\subsection{Maximally supersymmetric solutions}
\indent

The maximally supersymmetric solutions of ${\cal L} = {\cal L}_0 + {\cal L}_1$ we consider here were obtained in \cite{Castro:2007hc}. Since the lagrangian we are dealing with only has 1-form gauge fields, the near horizon background (\ref{senforms}) only allows for electrically charged solutions: to each vector multiplet corresponds an electrical charge $q_I$.

For our purposes it is enough to know the near horizon geometry, which is of the form (\ref{sen}) in $d=5,$ with $v_2= 4 v_1,$ and the field equations. For these maximally supersymmetric solutions, the $D$ field equation (\ref{fed}) can be written, near the horizon, as
\be
{\cal M} = 1 + \frac{1}{12 v_2} c_{I} M^I.
\ee

With this information one can compute the black hole entropy through Sen's formalism, which in this case gives $S= \frac{\pi^2}{4 G} v_1 v_2^{\frac{3}{2}} \left(F^I_{tr} \frac{\partial \mathcal{L}_H}{\partial F^I_{tr}} - \mathcal{L}_H \right),$ $\mathcal{L}_H$ being the lagrangian evaluated in the near horizon geometry. The final result is (for details see \cite{Castro:2007hc})
\be
S= \frac{\pi^2}{2 G} \mathcal{M} v_2^{\frac{3}{2}} =\mathcal{M} \frac{A_H}{4 G}, \label{larent}
\ee
$A_H$ being the horizon area.

One should in principle express this entropy in terms of the charges: for every choice of $\mathcal{M}$ (i.e for each choice of the set of numbers $c_{IJK}$ in (\ref{prep}) - or correspondingly for every choice of Calabi-Yau compactification manifold), that can be achieved by solving for $M^I$ the equation
\be \label{mbareq}
\frac{v_2}{2} c_{IJK} M^J M^K = q_I + \frac{c_{I}}{8}.
\ee
For our purposes it is enough to leave the entropy in the form (\ref{larent}), in terms of parameters of the near horizon metric.

As we previously mentioned, after elimination of the auxiliary fields (at least at the two derivative level) the lagrangian gets an overall factor of $\mathcal{M}:$ $\frac{\mathcal{M}}{16 \pi G} \R + \ldots$ This way we can redefine Newton's constant as $G_5 = G/\mathcal{M},$ in order to have the proper normalization of the lagrangian (even in the presence of the higher derivative corrections). With this interpretation of $G/\mathcal{M},$ we see that the entropy obtained in (\ref{larent}) verifies $S= \frac{A_H}{4 G_5}$. This means that, for these supersymmetric black holes with string-theoretical higher derivative corrections and $AdS_2 \otimes S^3$ near horizon geometry, from (\ref{seccaoext}) one has the simple relation between the entropy and the absorption cross section
\be
\sigma = 4 G_5 S.
\ee

\subsection{Heterotic string on $T^4\otimes S^1$}
\label{m123}
\indent

We now consider non-BPS solutions by taking a simple model with $I=1,2,3$ and prepotential
\be \label{stupp}
\mathcal{M} = M^{1} M^{2} M^{3} \;,
\ee
which is obtained taking M-theory compactified on a six-torus $T^6$, or equivalently \cite{Ferrara:1996hh, Kar:1995jx} heterotic string compactified on $T^4\otimes S^1$ winding around $S^1$ at tree level in the string coupling $g_S$.

The $D$ field equation (\ref{fed}) is equivalent to an algebraic on-shell condition for ${\cal M}$ (${\cal M}= 1$ at the two derivative level), implying that one has actually only two independent moduli. But, as it was shown in \cite{Kar:1995jx}, this compactification requires an additional 2-form $B$, with the corresponding strength $H=dB$, which can be obtained through a Poincar\'{e} duality transformation on the two-form gauge field $F^1=dA^1$.

The two derivative lagrangian ${\cal L}_0$ given by (\ref{l0}) with the prepotential (\ref{stupp}) admits three-charge black hole solutions (electric charges $q_2, q_3$ and magnetic charge $q_1$) whose near horizon geometry is of the form (\ref{sen}) in $d=5,$ with $v_2= 4 v_1 = \left|q_1q_2q_3\right|^{1/3}$ (in the Einstein frame). The black hole horizon area is given by $A_H= 2\pi^2 v_2^{3/2}$, from which we get the entropy
\be \label{stu0ent}
S = \frac{\pi^2}{2G} \sqrt{\left|q_1q_2q_3\right|}.
\ee

Article \cite{Cvitan:2007en} considers two kinds of higher derivative corrections (quadratic in the Riemann tensor) to these three-charge solutions: the full supersymmetric correction, and just the bosonic Gauss--Bonnet term. In the first case the corresponding bosonic lagrangian ${\cal L}_1$ is given by (\ref{l1cs}), while in the second case it is given by
\be
\frac{16 \pi G}{e} {\cal L}_1= \frac{c_{I} M^I}{192} Y_{\rm GB}, \label{l1gb}
\ee
$Y_{\rm GB}$ being given by (\ref{gb}).

In both cases, there are many different configurations of the coefficients $c_{I}$. For each of these configurations, the corrected black hole near horizon geometry has different branches, depending on the values of the charges one takes; for details, see \cite{Cvitan:2007en}.

In the presence of the full supersymmetric correction (\ref{l1cs}), the relation $v_2= 4 v_1$ for the near horizon geometry (\ref{sen}) is preserved. We take the following particular configuration of coefficients $c_{I}$:
\be
c_1 \equiv 24 \zeta >0, \, \, c_2=c_3 =0.
\ee
This set of coefficients, with $\zeta=1$, appears when we consider the heterotic string effective action on $T^4\otimes S^1$. We leave $\zeta$ explicit in order to remind that we are working perturbatively in $\a$ or, equivalently, in $\zeta$.

Just as an example, we focus on the branch of solutions such that $q_1 >\zeta/3, \, \, q_2>0, \, \, q_3 <0$. For the near horizon geometry we have $v_2 = \left|\frac{\left(q_1 + \zeta/3 \right)^2 q_2q_3}{q_1 - \zeta/3}\right|^{1/3}$, from which we get for the horizon area/absorption cross section
\be
A_H= \sigma = 2\pi^2 \left|\frac{\left(q_1 + \zeta/3 \right)^2 q_2q_3}{q_1 - \zeta/3}\right|^{1/2} \approx 2\pi^2 \sqrt{\left|q_1q_2q_3\right|} \left(1+\frac{\zeta}{2 q_1}\right).
\ee
The corrected entropy for these three-charge non-BPS black holes is of the form
\be \label{stu1ent}
S = \frac{\pi^2}{2G} \sqrt{\left|\left(q_1-\frac{\zeta}{3}\right)q_2q_3\right|} \approx \frac{\pi^2}{2G} \sqrt{\left|q_1q_2q_3\right|} \left(1-\frac{\zeta}{6 q_1}\right).
\ee
As one can see, the relation $\sigma = 4 G S$ is not verified to first order in $\zeta$.

In the presence of the Gauss--Bonnet correction, the relation between $v_1$ and $v_2$ is changed to $v_2= 4 v_1 - \frac{1}{2}$. There are also many different configurations of the second Chern class coefficients $c_{I}$ in (\ref{l1gb}), each of them also having different branches of solutions depending on the charges. We refer the reader to \cite{Cvitan:2007en} for a detailed analysis of all the branches of solutions corresponding to the different configurations. To our purposes, it is enough to state that, like for the full supersymmetric correction (\ref{l1cs}), the relation $S=\frac{A_H}{4 G}$ is not valid in the presence of this correction for the three-charge non-BPS black holes.

\section{Scattering of stringy scalars by 1/4 BPS $\mathcal{N}=4$ black holes in $d=4,5$}
\label{murthy}
\indent

Besides the examples we considered, in article \cite{Kuperstein:2010ka} the authors analyzed solutions of heterotic string theory with leading $\a$ corrections, corresponding to the low energy effective action (in the string frame) $S=S_0+S_1$:
\bea
S_0 &=& \frac{1}{16 \pi G_{10}}
\int \sqrt{- g} e^{-2\phi} \left[ \R+4(\nabla \phi )^2 -\frac{1}{2} \left\vert H_{3} \right\vert^2
 \right] d^{10}x, \nonumber \\
S_1 &=& \frac{1}{16 \pi G_{10}} \frac{\a}{8}
\int \sqrt{- g} e^{-2\phi} \left[ \R_{\mu\nu\rho\sigma} \R^{\mu\nu\rho\sigma} -\frac{1}{2} \R_{\mu\nu\rho\sigma} H_{\lambda}^{~\mu\nu} H^{\lambda\rho\sigma} \right. \nonumber \\
&-& \left. \frac{1}{8} H_{\kappa}^{~\mu\nu} H_{\lambda\mu\nu} H^{\kappa\rho\sigma} H^{\lambda}_{~\rho\sigma} + \frac{1}{24} H_{\kappa\lambda\mu} H^{\kappa}_{~\rho\sigma} H_{\nu}^{~\lambda\rho} H^{\nu\mu\sigma}\right] d^{10}x. \label{mea}
\eea
$H_{(3)} = d B_{(2)} + 3 \alpha^\prime \Omega_{(3)}$ is the NS-NS form, $\Omega_{(3)}$ being the gravitational Chern-Simons 3-form.

The solutions considered were 1/4 BPS $\mathcal{N}=4$ black holes in $d=5$ and $d=4$. On the first case, they corresponded to three charge solutions, with charges $n, w, W$, obtained by compactifying $S=S_0+S_1$ on a circle $S^1$ of length $2\pi R$ and a torus $T^4$ of four-volume $(2\pi)^4 V$. On the second case an extra compactification on a circle $\tilde{S}^1$ of radius $\tilde{R}$ was taken, and an extra charge $N$ was obtained, leaving us with a four charge solution with charges $n, w, N, W$ exactly analogous to those of the solution in section \ref{senglc}.

The near horizon metric of those solutions is as usual of the form (\ref{sen}), with
\bea
d&=&4: \,\, v_1=4\,N\,W+\frac{2\,\lambda}{G_4},\,
v_2=4\,N\,W+\frac{6\,\lambda}{G_4},\,
u_s=\sqrt{\frac{n\,w}{N\,W}}\left(1-\frac{1}{N\,W}\frac{\lambda}{G_4}\right), \nonumber \\
d&=&5: \,\, v_1=4\,W+\frac{2\,\lambda}{G_5},\,
v_2=4\,W+\frac{6\,\lambda}{G_5},\,
u_s=\sqrt{\frac{n\,w}{W}}\left(1-\frac{\lambda}{W\,G_5}\right).
\eea
Once again $\lambda$ has the same meaning as in (\ref{itef}). We could have written the previous quantities in terms of $\a$, but we chose to keep the same convention of section \ref{senglc}.
From these values, one can compute the black hole horizon area in the Einstein frame, obtaining
\bea
d&=&4: A_H= 8 \pi\sqrt{n\,w\,N\,W} \left(1-\frac{3}{4\,N\,W} \frac{\lambda}{G_4}\right), \nonumber \\
d&=&5: A_H= 8 \pi\sqrt{n\,w\,W} \left(1-\frac{3}{4\,W}\frac{\lambda}{G_5}\right). \label{areamurthy}
\eea

Differently than the other cases we have previously considered, for these 1/4 BPS black holes the corresponding massless scalar fields whose absorption cross section was computed in article \cite{Kuperstein:2010ka} were compactification moduli intrinsic to string theory (on that concrete case, compactified components of the metric). At the two derivative level, the action of any of these fields is the same action of a minimally coupled test field; therefore, according to the general result from \cite{dgm96}, at this level the low frequency limit of the cross section should be given by the horizon area. But to these fields one should consider $\a$ corrections to the cross section, coming not only from the black hole solutions (as we did in the previous cases) but also from the actions of the fields themselves, as they originate from string theory. These fields are not part of the black hole solution (which is the part of the metric which is not compactified): they vanish on-shell, and we are considering their fluctuations. In this sense they are test fields, as we defined in  section \ref{int}, but they have the same stringy origin as the black hole solution. They have their own stringy action with higher derivative corrections; therefore, they are not minimally coupled. This is the main difference from the cases we were previously considering.

In order to compute the cross section with all these $\a$ corrections, the authors of \cite{Kuperstein:2010ka} used the pole method, which was presented originally in \cite{Paulos:2009yk}. The cross section for a scalar field scattering off a black hole can be related, through the optical theorem, to the imaginary part of the forward scattering amplitude, which is the low frequency limit of a certain retarded Green's function. In such limit, for spherically symmetrical backgrounds, such Green's function can always be computed by an effective coupling in the higher derivative lagrangian evaluated at the black hole horizon. Such lagrangian has a simple pole in the radial coordinate, the residue of which corresponding to the desired Green's function. This is the essence of the pole method, which was presented originally in \cite{Paulos:2009yk}. The authors of \cite{Kuperstein:2010ka} used this method in order to compute the cross section with all the $\a$ corrections for the fields and black holes we mentioned, having obtained a result which was reminiscent of Wald's entropy formula, given by:
\bea
d&=&4: \sigma= 8 \pi G_4\sqrt{n\,w\,N\,W} \left(1+ \frac{1}{N\,W}\frac{\lambda}{G_4}\right), \nonumber \\
d&=&5: \sigma= 8 \pi G_5\sqrt{n\,w\,W} \left(1+\frac{3}{4\,W}\frac{\lambda}{G_5}\right). \label{entmurthy}
\eea
Indeed, for these stringy test fields with higher derivatives the agreement $\sigma= 4 G S$ between the low frequency absorption cross section and the black hole entropy was obtained to first order in $\a:$ up to the factor $4 G$, the expressions in (\ref{entmurthy}) correspond to the black hole entropy.

If we had considered minimally coupled test scalar fields, as before, their low frequency absorption cross section would be given, from (\ref{seccaoext}), by (\ref{areamurthy}). Once again it would not match the entropy up to a factor $4 G$, given by (\ref{entmurthy}).

\section{Discussion}
\label{discussion}
\indent

\subsection{Summary of the previous results}
\indent

We were interested in figuring out if and when, for extremal black holes, the higher derivative corrections to the entropy (given by Wald's formula or equivalently by Sen's formalism) leave it equal to one quarter of the low frequency absorption cross section (which, for minimally coupled test scalar fields, is given by the horizon area) or not. We have considered different cases (all with higher derivative corrections quadratic in the Riemann tensor), and we have obtained different answers.

In section \ref{emdgb}, we considered Einstein-Maxwell-dilaton black holes with Gauss-Bonnet corrections, with electric and magnetic charges (in $d=4$) and with just electric charges (in generic $d$ dimensions). For the first case, the corrections to the two quantities were always different. In the second case, in the absence of magnetic charges, the black holes became small. Their entropy was always proportional to the horizon area, but the proportionality factor was dimension-dependent and different than $1/4$.

For the extremal Reissner--Nordstr\"om solution in $d=4$ we considered in section \ref{itz}, the corrections to the two quantities were always different.

For the supersymmetric black holes in ${\cal N}=2$, $d=4$ supergravity considered in section \ref{glc}, the corrections to the two quantities were also always different, according to (\ref{ModIndEntropyFormula}). The same can be said about the four dimensional black holes with nonsupersymmetric Gauss-Bonnet corrections considered in section \ref{senglc}.

In section \ref{d5}, we considered black holes in ${\cal N}=2$, $d=5$ supergravity. We verified there was an agreement $\sigma= 4 G S$ for maximally supersymmetric purely electrical black holes, but that agreement did not exist in the presence of magnetic charges, for nonsupersymmetric black holes.

Finally in section \ref{murthy} we considered 1/4 BPS $\mathcal{N}=4$ black holes in $d=4,5$. The agreement $\sigma= 4 G S$ existed, but the test scalar fields had a stringy origin and were not minimally coupled. Their cross section was computed in a different way, through the pole method, and was not equal to the horizon area.

\subsection{Dependence on supersymmetry}
\indent

Because of this previously known example of 1/4 BPS $\mathcal{N}=4$ black holes in $d=4,5$ verifying $\sigma= 4 G S$ from \cite{Kuperstein:2010ka} mentioned in section \ref{murthy}, we may be led to the hypothesis that such relation being verified in the presence of $\a$ corrections by some black hole solution may be related to the amount of supersymmetry preserved by such solution. This hypothesis seemed to be confirmed by the examples considered in section \ref{d5}, where supersymmetric black holes preserved the relation $\sigma= 4 G S$ in the presence of $\a$ corrections, while nonsupersymmetric black holes did not.

But the examples from section \ref{glc} show clearly that the preservation of some supersymmetry does not imply such relation between $\sigma$ and $S$ in the presence of $\a$ corrections.

\subsection{Dependence on the charges}
\label{dischar}
\indent

The 1/4 BPS $\mathcal{N}=4$ black holes in $d=4,5$ verifying $\sigma= 4 G S$ from section \ref{murthy} have both electric and magnetic charges. In section \ref{d5} we found that the relation $\sigma= 4 G S$ was verified by maximally supersymmetric electrically charged black holes in $d=5$. We have previously concluded that the validity of such relation in the presence of higher order corrections did not depend on the amount of supersymmetry eventually preserved by the corresponding black holes. Could it be that it would always be valid in the absence of magnetic charges?

For the extremal string-corrected Reissner--Nordstr\"om solution in $d=4$ we considered in section \ref{itz}, the electric and magnetic charges are not independent: they both have to be present at the solution, and cannot simply be set to zero. Even if we express these quantities just in terms of the electric charge, the magnetic charge is also present. Since for this solution the corrections to $\sigma$ and $S$ are different, concerning this question this solution is not conclusive.

In section \ref{emdgb} we found that, concerning Einstein-Maxwell-dilaton black holes with Gauss-Bonnet corrections, for electrically charged black holes $\sigma$ was proportional to $S$ although in this case the proportionality factor was not necessarily $4 G$, while in the presence of magnetic charges there was no proportionality relation between $\sigma$ and $S$. The same can be concluded when setting the magnetic charges to 0 for the black holes of sections  \ref{senglc} and \ref{d5}.

The absence of magnetic charges corresponds to small black holes; in this case, the horizon area is proportional to $\sqrt{\a}$ and so is the entropy. Therefore small black holes with leading higher derivative corrections must have their area proportional to their entropy. For small black holes those leading corrections act as the leading terms in the action; in order to perform an analysis similar to the one we have been developing for these black holes, we would have to consider terms of even higher order, to see how would they affect their metric and entropy. That analysis is out of the scope of this article.

Considering the supersymmetric black holes in ${\cal N}=2$, $d=4$ supergravity from section \ref{glc}, the correction term to the entropy in (\ref{ModIndEntropyFormula}) is independent (at least explicitly) of the magnetic charges, which means in general that if we set these charges to 0, such correction term will keep being nonzero. Therefore the eventual validity of the relation $\sigma= 4 G S$ does not depend in general on the type of black hole charges.

\subsection{Dependence on the type and nature of the higher order correction}
\indent

The black holes we considered in section \ref{emdgb} were the only ones which did not have a supergravity/stringy origin. For all the other ones, we should distinguish between those which were solutions of a complete supersymmetric effective action, to first order in $\a$ (or its equivalent, after compactification to $d=4, 5$), and those which were solutions of classical Einstein gravity with a higher order correction term, not representing a full action to first order in $\a$, at least from the string theory point of view. This is the case of the four dimensional dyonic black holes in heterotic string theory we considered in section \ref{senglc}. These black holes (and their corresponding near horizon configurations (\ref{senv1v2}), (\ref{sendil}), (\ref{senf})) do not result from a complete supersymmetric action, but rather just from the Gauss--Bonnet term (\ref{sengb}): in \cite{Sen:2005iz} Sen observes that this higher derivative correction is enough to obtain the correct entropy (for a discussion on this point see \cite{Butter:2013lta}); yet, it may not be enough to obtain the correct $\a$-corrected low frequency absorption cross section, as the remaining terms in the action may change the near horizon geometries. A full supersymmetrization of this solution, like the one performed in \cite{Castro:2007hc}, may be necessary. But the full set of higher order terms in ${\cal N}=4$ supergravity is not known.

In article \cite{Cvitan:2007en} two kinds of higher derivative corrections to ${\cal N}=2$, $d=5$ supergravity were considered: the full supersymmetric correction, and just the bosonic Gauss--Bonnet term. As we mentioned in section \ref{m123}, the results were similar in both cases: the relation $\sigma= 4 G S$ is not valid in general.

\subsection{Distinction between external and stringy scalar fields}
\label{intrinsic}
\indent

From the \emph{limited} number of cases we have analyzed, we conclude that a full supersymmetric action and some supersymmetry preserved by the corresponding black hole solution \emph{may} be necessary but are definitely \emph{not} sufficient conditions for the relation $\sigma= 4 G S$ to be preserved in the presence of higher order corrections.

Clearly on dimensional grounds the absorption cross section must be proportional to the black hole entropy; the issue is if the proportionality constant depends or not on $\a$. But in the presence of higher derivative terms the metric, and hence the horizon area, are subject to field redefinition ambiguities (namely field redefinitions used to cancel higher order dilaton terms that appear after the dilaton dependent conformal transformation passing from string to Einstein frame).

As we have shown in \cite{Moura:2011rr}, for spherically symmetric $\a$-corrected $d$-dimensional black holes, up to a constant the only quantity which, classically (in the $\a \rightarrow 0$ limit), is proportional to the black hole horizon area and is invariant under metric frame transformations is the Wald entropy: the horizon area is not invariant under such transformations. This is not surprising: except for section \ref{murthy}, for the black hole solutions we have considered, the minimally coupled scalar fields of which we were computing the absorption cross section obviously did not have any stringy origin; there was no $\beta$ function from the nonlinear $\sigma-$model associated to them. Therefore the absorption cross section of these scalar fields, given by the horizon area, does not need to be invariant under metric frame transformations. On the other hand, for such black holes the black hole metric by itself is a solution of string theory, and therefore its entropy should be invariant under such frame transformations. We recall that the entropy of the black holes we considered in section \ref{senglc} was indeed invariant under such frame transformations, as we noticed there.

The example of the BPS $\mathcal{N}=4$ black holes in $d=4,5$ from section \ref{murthy} is different from the other cases we have mentioned. For these black holes the corresponding scalar fields whose absorption cross section was computed in \cite{Kuperstein:2010ka} were not minimally coupled external fields, as in the other examples: they were compactification moduli intrinsic to string theory (on that concrete case, compactified components of the metric). Their absorption cross section was computed using the pole method, which requires an action for the corresponding field to exist. Such fields were part of the action of the underlying theory, of which the corresponding black hole was a solution; that was not the case of the scalar fields of the other cases, which were external and introduced by hand.

One must therefore distinguish between two cases: scattering off an extremal black hole of a minimally coupled external  scalar field and of a higher derivative scalar field (obeying a higher order field equation).

On the first case, as we have seen, even in the presence of higher derivative corrections to the black hole metric there is a universal limit for the low frequency absorption cross section, given from (\ref{seccaoext}) by the area of the black hole horizon, at least for spherically symmetrical solutions.

On the second case, one is not on the conditions of the result of \cite{dgm96} or the derivation (\ref{seccaoext}): in both cases one was assuming an external test scalar field, with a second order field equation. One must therefore proceed differently in order to compute its absorption cross section. In article \cite{Kuperstein:2010ka} the pole method was used for this computation, and the result $\sigma= 4 G S$ was obtained, to first order in $\a$, for specific black hole solutions. Now, the question is: how general could this result be?

Because of the stringy origin of the scattered scalar field, at least in string theory its cross section should be independent of the chosen metric frame. As we have shown in \cite{Moura:2011rr}, up to a constant the only quantity classically proportional to the horizon area which is independent of the metric frame is the black hole entropy. Therefore, for a scalar field of this type its absorption cross section should naturally be proportional to the black hole entropy. Now it is well known that, in the presence of higher derivative corrections, the black hole Wald entropy is in general not proportional to the black hole horizon area. This means that for a scalar field of this type the low frequency absorption cross section is \emph{not} its area. We checked this explicitly, to first order in $\a$, for the solutions we analyzed in section \ref{murthy}. This does not contradict the result of \cite{dgm96} nor the derivation (\ref{seccaoext}), as in both these cases one was assuming an external test scalar field, with a second order field equation. Naturally, in the $\a \rightarrow 0$ limit, $S=A_H/4G$ and we would keep having $\sigma=A_H$ according to \cite{dgm96} or (\ref{seccaoext}).

Therefore, the absorption cross section for black holes in string theory depends on the nature of the scattered scalar field: the result for external minimally coupled scalar fields is different than the one for intrinsically stringy, nonminimally coupled scalar fields. Of course for some particular black hole solutions the two cases may give the same result, which corresponds to the Wald entropy remaining equal to $A_H/4G$ to first order in $\a$: that was the case, for instance, of the maximally supersymmetric electrically charged black holes in $d=5$ from section \ref{d5}. This coincidence may happen for this and other solutions but, in general, the two results will be different.

\section{Conclusions and future work}
\indent

In this article we have obtained a general formula for the low frequency absorption cross section of minimally coupled test scalar fields by extremal spherically symmetric $d$-dimensional black holes, including higher derivative corrections to arbitrary order, as long as those corrections preserve the near horizon $AdS_2 \otimes S^{d-2}$ geometry. This is the case of $\a$ corrections in string theory. This formula is exactly the same as the classical one: the low frequency absorption cross section $\sigma$ is equal to the horizon area $A_H$.

Classically the relations $A_H = \sigma = 4~G~S$ are valid. The question then amounts to figuring out if and when, for these black holes, the $\a$ corrections to the black hole entropy $S$ (given by Wald's formula or equivalently by Sen's formalism) leave it equal to $A_H/4G.$ We have considered different kinds of extremal black hole solutions, supersymmetric and nonsupersymmetric, with electric and with magnetic charges, with higher order corrections of different nature. We have obtained enough counterexamples to state that the agreement $\sigma= 4 G S$ is in general not verified for scattering of minimally coupled scalars by black holes with higher order corrections.

All the cases we met in which such agreement was verified correspond to supersymmetric black holes. But for two of these cases, from section \ref{murthy}, the scattered scalar field was not a minimally coupled field: it also had higher derivative corrections, and a stringy nature. The corresponding cross section was therefore computed through a different procedure: the pole method. This suggests that the result for the cross section of intrinsically stringy scalar fields is indeed different than the one for minimally coupled external fields.

In section \ref{intrinsic} we argued that the absorption cross section of intrinsically stringy scalar fields should be independent of the metric frame. Based on that argument and in our proof from \cite{Moura:2011rr} that for a class of stringy black holes with $\a$ corrections the only quantity classically proportional to the horizon area which is invariant under metric frame transformations is, up to a constant, the Wald entropy, we concluded that, in string theory, the absorption cross section of such scalar fields should in general be given by $\sigma= 4 G S$.

The observations we have made in this article were formulated by considering just some examples of black holes in $d=4,5$ with higher order corrections, but few of these examples exist in the literature. Fortunately those which we have considered were representative enough to allow for some conclusions to be taken. Nonetheless, it would be certainly pleasant to have more examples showing explicitly the agreement $\sigma= 4 G S$ in the presence of higher order corrections for stringy scalar fields. It would be even better to have a formal proof that such agreement always exists for those cases: what we have are hints and arguments, suggested by the analysis of the few known positive cases.

Also, the argument that the absorption cross section should be independent of the metric frame we presented in section \ref{intrinsic} is clearly only valid in the context of string theory. It would certainly also be nice if one could extend such study for generic black holes with higher derivative corrections, either by providing counterexamples or by giving a formal proof of the agreement, and not only examples and arguments based on string theory, as we did here. Nonetheless, the absorption cross section is computed, through the pole method, using the higher derivative terms of the scalar field. The black hole entropy is computed, through Wald's formula, using the higher powers of the Riemann tensor, i.e. other type of higher derivative terms. The agreement $\sigma= 4 G S$ requires that the coefficients of these higher order terms are not arbitrary (at least, not all). This is another reason to suspect that such agreement could only be valid in string theories: in these theories, after a specific background is chosen, the coefficients in the effective action should not be arbitrary. It is possible that, for a specific black hole background, only a stringy test field has the right coefficients in its effective action in order to match its absorption cross section to the black hole entropy.

These are a few topics in this subject that can be considered in future works.

\section*{Acknowledgments}
The author wishes to acknowledge very useful discussions with Gabriel Lopes Cardoso and Predrag Dominis Prester. This work has been supported by the European Union's Horizon 2020 Programme and by Funda\c c\~ao para a Ci\^encia e a Tecnologia through projects LIP/50007 and CERN/FP/123609/2011.


\bibliographystyle{plain}

\end{document}